\documentclass[english,a4paper,prl,twocolumn,amsmath,amssymb,longbibliography,superscriptaddress]{revtex4-1}
\usepackage{bm}
\usepackage{graphicx,xcolor}
\usepackage[colorlinks,linkcolor=magenta,citecolor=magenta]{hyperref}

\newcommand{\dd}{\mathrm{d}}

\newcommand{\pd}[2]{\frac{\partial #1}{\partial #2}}

\newcommand{\Int}[1]{\int\dd #1\;}
\newcommand{\IInt}[3]{\int_{#2}^{#3}\dd #1\;}

\renewcommand{\vec}[1]{\mathbf #1}

\newcommand{\al}{\alpha}

\newcommand{\eps}{\varepsilon}
\newcommand{\kap}{\kappa}

\newcommand{\vhi}{\varphi}
\newcommand{\sig}{\sigma}
\newcommand{\om}{\omega}

\newcommand{\id}{\mathbf 1}
\newcommand{\im}{\text i}
\newcommand{\x}{\vec r}

\begin{document}

\title{Collective Hall current in chiral active fluids: Coupling of phase and mass transport through traveling bands}

\author{Frank Siebers}
\affiliation{Institut f\"ur Physik, Johannes Gutenberg-Universit\"at Mainz, Staudingerweg 7-9, 55128 Mainz, Germany}
\author{Robin Bebon}
\author{Ashreya Jayaram}
\author{Thomas Speck}
\affiliation{Institute for Theoretical Physics IV, University of Stuttgart, Heisenbergstr.~3, 70569 Stuttgart, Germany}

\begin{abstract}
  Active fluids composed of constituents that are constantly driven away from thermal equilibrium can support spontaneous currents and can be engineered to have unconventional transport properties. Here we report the emergence of (meta-)stable traveling bands in computer simulations of aligning circle swimmers. These bands are different from polar flocks and we show that they can be understood as non-dispersive soliton solutions of the underlying non-linear hydrodynamic equations with constant celerity (phase propagation speed) that is much larger than the propulsion speed. In contrast to solitons in passive media, these bands can induce a bulk particle current with a component perpendicular to the propagation direction, thus constituting a collective Hall (or Magnus) effect. Traveling bands require sufficiently small orbits and undergo a discontinuous transition into a synchronized state with transient polar clusters for large orbital radii.
\end{abstract}

\maketitle

%% ---- introduction ----

Motile active matter composed of self-propelled constituents exhibits a plethora of collective states that have attracted copious interest over the last decade~\cite{ramaswamy10,ginelli10,vicsek12,fily12,buttinoni13}. The statistical physics of active matter is rapidly evolving into a comprehensive theoretical framework to address the emergence of spatiotemporal structures away from thermal equilibrium, with a particular focus on living matter~\cite{almonacid15,shelley16,needleman17}. Minimal models have been instrumental in exposing physical principles underlying these emergent states. For example, the seminal Vicsek model~\cite{vicsek95} of idealized aligning spin particles explains the formation of polarized flocks via a phase transition~\cite{chate08,caussin14,solon15}.

In contrast to active particles that propel linearly, \emph{chiral} active particles---including bacteria~\cite{diluzio05,lauga06}, sperm cells moving near boundaries~\cite{kaupp03,friedrich07}, and anisotropic colloidal microswimmers~\cite{kummel13,zhang20}---additionally self-rotate about a given axis resulting in circular (or helical) motion that is, in general, perturbed by fluctuations~\cite{lowen16,liebchen17,levis18,lei19,liebchen22}. Chirality on the individual level leads to the emergence of a wealth of collective behavior: rotating dynamical crystallites~\cite{huang20}, vortex arrays~\cite{kaiser13}, chimera states~\cite{kruk20}, hyperuniformity~\cite{lei19}, stereoselectivity~\cite{arora21}, hydrodynamic synchronization~\cite{samatas23}, Kelvin waves~\cite{poggioli23a}, and ``odd'' viscosity~\cite{banerjee17,liao19,lou22,poggioli23}. Recently, the susceptibility to chirality disorder has been studied~\cite{ventejou21,siebers23}. In the limit of vanishing linear propulsion, active ``spinners" undergo phase separation, self-assemble into lattices, synchronize and generate vortical/turbulent flows~\cite{nguyen14,yeo15,sabrina15,vanzuiden16,gorce19,goto15,kokot17,shen19}. Interestingly, materials composed of spinners can sustain topologically protected edge modes that propagate robustly through the sample~\cite{vanzuiden16,soni19}.

In the absence of persistent rotation, active particles phase-separate into dense and dilute regions at sufficiently large global densities and propulsion speeds~\cite{buttinoni13,fily12,cates15}. Such coexistence of dense and dilute regions has also been reported in several flavours of aligning active particles~\cite{bar20,chate20}. Specifically, in the Vicsek model particles aggregate into dense bands that propagate through a dilute background perpendicular to the interface~\cite{chate08a,caussin14,solon15}. Polar rods, on the other hand, assemble into bands that travel parallel to the interface~\cite{ginelli10,farrell12,weitz15,jayaram20}. Nematically ordered bands that dynamically merge and disintegrate have been observed in active nematics~\cite{chate06,shi14}.

%% ---- system ----

\begin{figure}[t]
  \centering
  \includegraphics{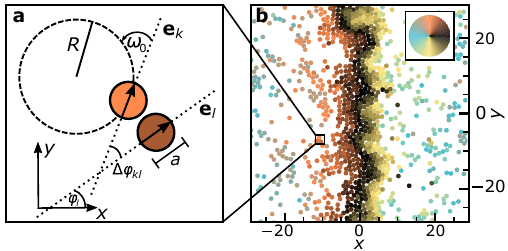}
  \caption{\textbf{Chiral active particles form traveling bands.} (a)~Sketch of two particles with diameter $a$ moving along their orientation vectors $\vec e_k$ with speed $v_0$. The orientations change with angular speed $\om_0$, which implies closed orbits with radius $R=v_0/|\om_0|$ for noiseless non-interacting particles. When coming within distance $|\x_l-\x_k|<\sig$, particles experience an additional alignment interaction depending on the difference $\Delta\vhi_{kl}=\vhi_l-\vhi_k$. (b)~Simulation snapshot of a traveling band (from left to right) in a square box for propulsion speed $v_0=60$ and angular speed $\om_0=6$. The color indicates the particle orientation (see color wheel).}
  \label{fig:sys}
\end{figure}

In this Letter, we demonstrate that aligning chiral active particles with identical angular speeds can form \emph{non-dispersive} traveling bands (in lieu of previously observed ``microflocks''~\cite{liebchen17}) and investigate their characteristics. To this end, we study a minimal model of $N$ particles moving in two dimensions that is related to both the (continuous-time) Vicsek model and the Kuramoto model~\cite{kuramoto75,acebron05}. The evolution of position $\x_k$ and orientation $\vhi_k$ of the $k$th particle is governed by the overdamped equations of motion
\begin{gather}
  \label{eq:r}
  \dot\x_k = v_0\vec e_k + \mu_0\vec F_k, \\
  \label{eq:vhi}
  \dot\vhi_k = \om_0 + \frac{\kap}{\pi\sig^2}\sum_{l\in n_k}\sin(\vhi_l-\vhi_k) + \sqrt{2/\tau}\eta_k,
\end{gather}
where $\eta_k$ is a random number drawn from a unit normal distribution. Particles have diameter $a$ and interact with their neighbors via aligning torques and short-ranged repulsive forces $\vec F_k$ modeling volume exclusion. As sketched in Fig.~\ref{fig:sys}(a), each particle has a unit orientation vector $\vec e_k=(\cos\vhi_k,\sin\vhi_k)^\top$ along which it is propelled with constant speed $v_0$. The orientations are affected by an angular speed $\om_0$ and undergo rotational diffusion with correlation time $\tau$. Hereinafter, we report numerical values for lengths in units of $a$ and times in units of $\tau$.

In the absence of diffusion, a free particle would move along a circular orbit of radius $R\equiv v_0/|\om_0|$. The $k$th and $l$th particles align with strength $\kap=40$ if $|\x_l-\x_k|<\sig$ with interaction radius $\sig=1.25$. Our model reduces to several other models as special cases: active Brownian particles ($\om_0=0$ and $\kap=0$, see Ref.~\citenum{lei19} for $\om_0\neq0$), the Vicsek model ($\om_0=0$ and $\vec F_k=0$), and fixing particle positions ($\dot\x_k=0$) leads to the noisy Kuramoto model of coupled oscillators with phases $\vhi_k$.

% ---- simulations I: definitions and phenomenon ----

We perform Brownian dynamics simulations of $N=1000$ particles in an $L_x\times L_y$ box with periodic boundary conditions (see SM~\cite{sm} for details). Unless stated otherwise, we set $L_x=L_y$ in the simulations. Throughout, we consider a global density $\bar\rho=N/(L_xL_y)=0.3$ corresponding to a packing fraction $\phi\simeq0.24$. At this density, we find that non-zero angular speeds $\om_0\neq 0$ inhibit motility-induced phase separation. Interestingly, we observe that for certain speeds $v_0$ and $\om_0$ a band of increased density might form spontaneously, cf.~Fig.~\ref{fig:sys}(b). Single particles continue to move along (almost) closed orbits of radius $R$ but get temporally bunched into a denser band. This band travels at a constant speed $u$ larger than the propulsion speed $v_0$. The propagation direction is in principle random but aligns with the edges of our finite simulation box.

% ---- theory I: perturbation theory ----

To understand the nature of these bands, we turn to the effective hydrodynamic equations for the local density $\rho(\x,t)$ and polarization $\vec p(\x,t)$ starting from Eqs.~\eqref{eq:r} and~\eqref{eq:vhi}. While the time-evolution of the density simply follows the continuity equation $\partial_t\rho+v_0\nabla\cdot \vec{p}=0$, the polarization is governed by (SM~\cite{sm})
\begin{equation}
  \partial_t\vec p = -\frac{v_0}{2}\nabla\rho + \om_0\vec R\cdot\vec p + \left(\frac{\kap}{2}\rho - \frac{1}{\tau} - \al|\vec p|^2\right)\vec p.
  \label{eq:p}
\end{equation}
Importantly, our approach predicts that the angular speed $\om_0$ is not renormalized (in contrast to Ref.~\citenum{sese-sansa22}). Moreover, the density-dependent reduction of the propulsion speed due to interactions is negligible (SM~\cite{sm}) and we continue to employ the bare speed $v_0$. Equation~\eqref{eq:p} contains $-\al|\vec p|^2\vec p$ as the lowest-order correction arising from interactions (and the closure relation for the nematic tensor) with effective coefficient $\al$. This term with $\al>0$ is the minimal ingredient to limit the growth of the polarization.

\begin{figure}[t]
  \centering
  \includegraphics{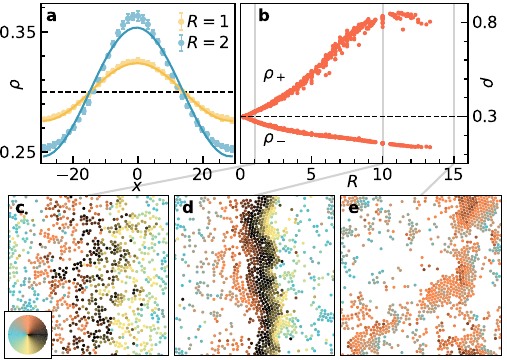}
  \caption{\textbf{Characterization of density profiles.} (a)~Numerical density profiles $\rho(x)$ in the comoving frame for small $R/L_x\ll 1$. The black dashed line indicates the global density $\bar\rho=0.3$ and the solid lines are fits to Eq.~\eqref{eq:rho0}. (b)~Band ($\rho_+$) and gas ($\rho_-$) densities as a function of $R$. Every point corresponds to an independent simulation. A total of 437 simulation runs are shown with $v_0\in[20,150]$ and $\om_0\in[3,100]$, demonstrating that the densities $\rho_\pm$ collapse when plotted as a function of orbital radius $R=v_0/|\omega_0|$. (c-e)~Snapshots of the system with (c)~$R=1$ (traveling band with local polarization progressing along the $x$-direction), (d)~$R=10$ (dense traveling band with mass transport), and (e)~$R=15$ (synchronized state: transient clusters with large global polarization magnitude). See also Supplemental Videos 2-4~\cite{sm}.}
  \label{fig:den}
\end{figure}

In steady state, a uniform density profile $\rho=\bar\rho$ implies vanishing polarization for $\chi\equiv\kap\bar\rho/2-\tau^{-1}<0$. For $\chi>0$, alignment overcomes rotational diffusion (in our simulations $\chi=5$) and the solution for uniform density is a rotating polarization $\vec p^{(\text s)}(t)=\sqrt{\chi/\al}(\cos\om_0t,\sin\om_0t)^\top$ with fixed magnitude. In the Kuramoto model, $\vec p^{(\text s)}$ describes the fully \emph{synchronized state}. Although simulation snapshots show transient ``microflocks'' (in agreement with~\cite{levis18}) the polarization does follow $\vec p^{(\text s)}$, cf. Fig.~\ref{fig:den}(c). The linear stability of this synchronized state is determined by a competition of $\kap$ (alignment) and $\chi$ (relaxation) independent of $v_0$ and $\om_0$, with the synchronized state becoming linearly unstable for both small and large alignment strength $\kap$ (SM~\cite{sm}). Our simulations are performed in the stable region, from which the system spontaneously nucleates the traveling bands. Since these bands are the result of rare events, we promote their formation through initializing the orientations of half of the particles according to $\vhi_k=-2\pi x_k/L_x$ while leaving the others at random orientations (see SM~\cite{sm} for details).

To investigate how these traveling planar bands can emerge, we align the propagation direction with the $x$-axis and change to a \emph{comoving frame} through $x\to x-ct$ with (yet unknown) celerity $c$. In this comoving frame, both density $\rho(x)$ and polarization $\vec p(x)$ only depend on $x$. The continuity equation, which then reads $-c\partial_x\rho=-v_0\partial_x p_x$, can be integrated to
\begin{equation}
  \rho(x) = \rho_0 + \eps p_x(x)
  \label{eq:rho}
\end{equation}
with $\eps\equiv v_0/c$. Here, $\rho_0$ is an integration constant determined by density conservation but for the following discussion $\rho_0=\bar\rho$ (SM~\cite{sm}).

For small $\eps\ll 1$, i.e., celerities $c\gg v_0$ much larger than the propulsion speed $v_0$, we perturbatively expand $\vec p=\vec p^{(0)}+\eps\vec p^{(1)}+\cdots$. To lowest order of $\eps$, we have to solve the differential equation
\begin{equation}
  -c\partial_x\vec p^{(0)} = \om_0\vec R\cdot\vec p^{(0)} + (\chi-\al|\vec p^{(0)}|^2)\vec p^{(0)},
  \label{eq:p0}
\end{equation}
which permits spatially periodic solutions
\begin{equation}
  \vec p^{(0)}(x;q) = \sqrt{\chi/\al}\left(\begin{array}{c}
    \cos qx \\ -\sin qx
  \end{array}\right).
  \label{eq:p0:sol}
\end{equation}
These solutions describe the uniform rotation of the polarization vector (still with fixed magnitude $\sqrt{\chi/\al}$) as we move along the $x$-axis, essentially unraveling the temporally periodic $\vec p^{(\text s)}$ into (comoving) space. A spatially varying polarization does, however, imply a non-uniform density through Eq.~\eqref{eq:rho}. Plugging Eq.~\eqref{eq:p0:sol} into Eq.~\eqref{eq:p0} fixes the celerity to $c(q)=\om_0/q$, i.e., to a linear dispersion relation between the external ``frequency'' $\om_0$ and the wave vector $q$. Note that in a finite system (as in the simulations), periodic boundaries enforce discrete wave vectors $q_n=2\pi n/L_x$.

% ---- simulations II: low \eps data ----

Although multiple stable bands can be observed in elongated simulation boxes (Supplemental Video 1~\cite{sm}), we now focus on a single band ($n=1$) in a square box. With the celerity $c=\om_0/q_1$ we find $\eps=2\pi R/L_x$ and, therefore, we first turn to simulations for which $R\ll L_x$. In Fig.~\ref{fig:den}(a), we plot density profiles in the comoving frame together with fits to 
\begin{equation}
  \rho(x) = \bar\rho + A\cos\frac{x}{2\pi L_x}
  \label{eq:rho0}
\end{equation}
predicted through Eq.~\eqref{eq:rho} together with Eq.~\eqref{eq:p0:sol}. The agreement is excellent. Here we fit the amplitude $A$, from which we determine the interaction coefficient $\al$. On increasing $R$, the density profile starts to deviate from a sinusoid and the quality of fits deteriorates. The profile becomes more localized with a dense band surrounded by a dilute gas [Fig.~\ref{fig:den}(d)]. Occasionally, bands break and enter the synchronized state (SM~\cite{sm}). We construct stationary density profiles in the comoving frame using only simulation data in the traveling band state. From fits of these density profiles, in Fig.~\ref{fig:den}(b) we plot the coexisting band and gas densities $\rho_\pm$. For small $R$, the difference $\rho_+-\rho_-\propto R$ increases with $R$ as predicted from $A\propto\eps\propto R$. Notably, simulations for different speeds $v_0$ and $\om_0$ collapse onto a master curve that only depends on the orbital radius $R$. For large orbits $R\gtrsim 13$ bands become unstable [Fig.~\ref{fig:den}(e)] and we return to this regime later.

% ---- theory II: currents ----

We now move to the next order of $\eps$, for which the evolution equation for the polarization in the comoving frame becomes
\begin{equation}
  -c \partial_x \vec p^{(1)} = \om_0 \vec R \cdot \vec p^{(1)} + \frac{\kap}{2} p_x^{(0)} \vec p^{(0)} - 2\al \left(\vec p^{(0)} \cdot \vec p^{(1)} \right) \vec p^{(0)}.
  \label{eq:p1}
\end{equation}
Using Eq.~\eqref{eq:p0:sol} as an ansatz but with a position-dependent magnitude, we obtain (see SM~\cite{sm} for details)
\begin{equation}
  \vec p^{(1)}(x;q) = \beta\left[ \chi \cos qx - \frac{\om_0}{2} \sin qx\right]\left(\begin{array}{c}
    \cos q x \\ -\sin q x
  \end{array}\right)
  \label{eq:p1:sol}
\end{equation}
with prefactor $\beta\equiv\kap\chi/[\al(4\chi^2+\om_0^2)]$. Importantly, we find the same linear dispersion relation $c(q)=\om_0/q$. Unlike Eq.~\eqref{eq:p0:sol}, components of the polarization at first order are no longer centered around $x=0$ and additionally lose their axi- ($p_x$) and point-symmetry ($p_y$).

These broken symmetries hint at an alignment imbalance between particles at the front and back of the band. While the direction of the polarization field has to rotate once within one period $\tau_1=2\pi/\om_0=L_x/c$, a change of its magnitude implies a displacement, i.e., particles do not (on average) return to their initial position. This imbalance thus prompts the onset of non-vanishing particle currents in the laboratory frame. Averaging over one period $\tau_1$, we find (SM~\cite{sm})
\begin{equation}
  J_x = v_0\frac{L_y}{L_x}\IInt{x}{0}{L_x} p_x(x), \quad
  J_y = v_0\IInt{x}{0}{L_x} p_y(x)
\end{equation}
for the currents through the respective cross sections calculating the integrals in the comoving frame. Clearly, the zeroth order solution $\vec p^{(0)}$ does not contribute due to its symmetry. In contrast, the first-order correction gives rise to directed mass transport with magnitudes
\begin{equation}
  J^{(1)}_x = \frac{1}{2}v_0\eps\beta\chi L_y, \quad J^{(1)}_y = \frac{1}{4}v_0\eps\beta\om_0L_x.
  \label{eq:cur}
\end{equation}
Note the quadratic scaling in the self-propulsion speed $v_0$ since $\eps=v_0/c$. Surprisingly, the directions of mass transport and wave propagation are neither parallel nor perpendicular. Moreover, $J_x$ is antisymmetric under reversal of the rotation direction $\om_0\to-\om_0$, i.e., $J_x(-\om_0) = -J_x(\om_0)$, while $J_y$ is invariant since the polarization wave propagates along $x$ by choice of the coordinate system.

% ---- simulations III: higher \eps, current ----

\begin{figure}[t]
  \centering
  \includegraphics{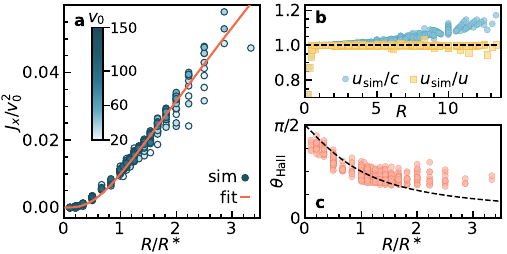}
  \caption{\textbf{Particle current couples to phase propagation.} (a)~Scaled particle current $J_x/v_0^2$ in propagation direction as function of the dimensionless orbital radius $R/R^\ast$. Circles correspond to simulation data (colored according to their self-propulsion speed $v_0$). The solid orange line is obtained by fitting the right-hand side of Eq.~\eqref{eq:cur_scaled} with parameter $\al$. (b)~The measured speed $u_\text{sim}$ normalized by the celerity $c$ (blue dots) and the travel speed $u=c+v_x$ including the current-induced speed $v_x$ (yellow squares). The dashed black line indicates unity, highlighting the deviation if the current contribution is ignored and the excellent agreement between prediction and simulation if current-induced speeds are accounted for. (c)~Measured Hall angles $\theta_\mathrm{Hall}$ as function of $R/R^\ast$. The dashed black line indicates the theoretical linear-order prediction $\theta_\text{Hall}=\tan^{-1}(R^\ast/R)$.}
  \label{fig:cur_speed}
\end{figure}

We can rewrite the current along the propagation direction
\begin{equation}
  \frac{J_x}{v_0^2} = \frac{\kap\pi}{8\al\chi}\frac{L_y}{L_x}\frac{(R/R^\ast)^3}{1+(R/R^\ast)^2}
  \label{eq:cur_scaled}
\end{equation}
with length $R^\ast\equiv v_0/(2\chi)=v_0\tau/(\kap\tau\bar\rho-2)$. Figure~\ref{fig:cur_speed}(a) shows the numerically determined currents $J_x$ as a function of $R/R^\ast$ together with a fit to Eq.~\eqref{eq:cur_scaled} yielding $\al\simeq159$ as the only fit parameter. While this prediction has been derived to linear order in $\eps$, it does capture the behavior even for larger orbits quite well.

Displacing particles on average during one period implies that the density profile (the band) has to travel with a larger speed $u>c$ while phases still propagate with celerity $c$. In the simulations, we directly measure the speed $u_\text{sim}$ from the position $x_0=u_\text{sim}t$ where the average orientation aligns with the propagation direction. In Fig.~\ref{fig:cur_speed}(b), the reduced speed $u_\text{sim}/c$ with celerity $c$ is plotted as a function of $R$. While for small $R$ both agree, for larger orbits the bands indeed move faster. To account for the current, we have to add the associated speed $v_x=J_x/(\bar\rho L_y)$ through a cross section with length $L_y$, which leads to the travel speed $u=c+v_x$ of the band. Figure~\ref{fig:cur_speed}(b) shows that $u_\text{sim}=u$ over the full range of orbits up to the point where bands break and the system enters the synchronized state. The deviations for very small $R$ can be attributed to the system entering an absorbing state~\cite{lei19}.

Note that the perpendicular current $J_y$ induces the speed $v_y=J_y/(\bar\rho L_x)$, but this current does not influence the travel speed $u$. Rearranging Eq.~\eqref{eq:cur}, we can express the ratio $v_y/v_x=\om_0/(2\chi)=R^\ast/R=\tan\theta_\text{Hall}$ of transversal to longitudinal speeds through a Hall angle $\theta_\text{Hall}$. In contrast to the Hall effect observed for a probe driven through a non-aligning chiral active bath~\cite{reichhardt19}, here the transverse current is an emergent collective phenomenon. In Fig.~\ref{fig:cur_speed}(c) we plot $\theta_\text{Hall}$ extracted from the simulations. While the collapse and overall quantitative agreement is less good, the data follows the trend predicted from the linear-order solution. The particle current is almost perpendicular ($\theta_\text{Hall}\approx\pi/2$) to the propagation direction for small $R$ and seems to saturate at $\theta_\text{Hall}\approx\pi/4$ in the limit of large $R$.

% ---- breaking of band ----

As already alluded to, for larger orbits the bands break and the system enters the synchronized state. To distinguish the two states and to identify transitions, we employ the global order parameter
\begin{equation}
  P(t) \equiv \frac{1}{N}\left|\sum_{k=1}^N\vec e_k(t)\right|
\end{equation}
measuring the magnitude of the average orientation in the system. This order parameter is close to unity in the synchronized state at all times and should approach zero for traveling bands since, for a spatially periodic polarization, all orientations are present and thus cancel. Figure~\ref{fig:break}(a) shows a time trace at $R=13.3$ demonstrating the sudden rupture of the band with the system subsequently entering the synchronized state. We also find (rare) transitions at much smaller $R$ (SM~\cite{sm}), which indicates that traveling bands are susceptible to fluctuations. These observations suggest a discontinuous transition from bands to the synchronized state, both of which are linearly stable against small fluctuations.

\begin{figure}[b!]
  \centering
  \includegraphics{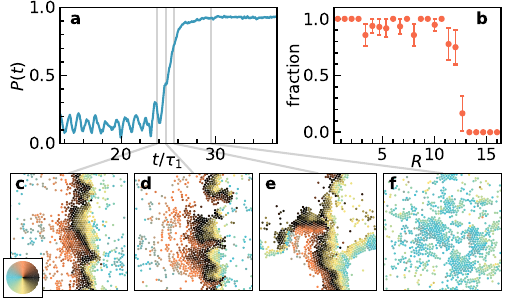}
  \caption{\textbf{Discontinuous transition to the synchronized state.} (a)~Global order parameter $P(t)$ as a function of scaled time $t/\tau_1$ for $R=13.3$ showing a sharp increase at $t\approx 25\tau_1$ (with $\tau_1=2\pi/\om_0$). (b)~Fraction of simulation runs (with initial orientations favoring traveling bands) that have survived for $t=45$ displaying a sharp transition at $R\approx 13$. Simulations for different $v_0$ and $\om_0$ are binned and error bars indicate the standard error. (c-f)~Corresponding snapshots showing the band's rupture.}
  \label{fig:break}
\end{figure}

We speculate that the mechanism leading to the destabilization of bands is indeed a density fluctuation: for a stable band particles need to constantly join the dense region from the front. As the average density directly in front of the propagating band decreases, instances in which not enough particles join uniformly along the interface become more likely. In this event, the band ruptures and, due to the non-zero angular speed, curls upwards, cf. Fig.~\ref{fig:break}(d). In Fig.~\ref{fig:break}(b) we plot the fraction of bands that have survived for the duration of the simulation ($t=45$). While there are some runs that reach the synchronized state even at small $R$ (presumably due to rare large fluctuations), there is a sharp transition to the synchronized state at $R\approx 13$ beyond which bands can no longer be stabilized.

%% ---- conclusions ----

To conclude, we have studied a minimal model for chiral active fluids composed of circle swimmers, for which we have uncovered the emergence of spontaneous waves and currents: within the linearly stable regime of the fully synchronized state traveling bands of increased density can form. These bands are controlled by the orbital radius $R$ of individual particles and they are characterized by two speeds different from their propulsion speed $v_0$: the first is the celerity $c$ imposed by the frequency $\om_0$ and the spatial extent $L_x$. This speed governs the non-dispersive propagation of particle orientations (phases). However, a non-uniform polarization displaces particles and bands have to travel with a second, different speed $u$. The displacements give rise to a particle current, and thus mass transport, which not only has a component along the propagation direction of the band but also a component perpendicular. It thus constitutes a \emph{collective} Hall effect absent on the level of individual particles. Importantly, currents are not externally forced (as for ``odd'' viscosity~\cite{liao19,lou22}) but emerge spontaneously due to spatiotemporal synchronization. Developing a minimal and transparent hydrodynamic theory based on a single fit parameter has allowed us to corroborate the numerical observations. Our results open new routes for the development of active (meta)fluids with programmable and topological flow states.

%% ---- acknowledgments ----

\begin{acknowledgments}
	We acknowledge funding by the Deutsche Forschungsgemeinschaft (DFG) within collaborative research center TRR 146 (Grant No.~404840447). Computations have been performed on the HPC cluster MOGON II.
\end{acknowledgments}

%% ---- bibliography ----

  
%% ---- supplemental ----

\onecolumngrid
\newcommand\supplement{%
       \setcounter{table}{0}
       \renewcommand{\thetable}{S\arabic{table}}
       \setcounter{figure}{0}
       \renewcommand{\thefigure}{S\arabic{figure}}
       \setcounter{equation}{0}
       \renewcommand{\theequation}{S\arabic{equation}}
       \setcounter{section}{0}
       \renewcommand{\thesection}{S\arabic{section}}
    }

\section{Simulation Details}

Simulations were performed using an in-house code, integrating the equations of motion [Eqs.~(1,2) of the main text] with a time step $\delta t=10^{-6}$ in a rectangular box $L_x \times L_y $ with periodic boundary conditions. If not stated otherwise, the following parameters remain unchanged in order to limit the parameter space and focus on the dominant $R$-dependence: number of particles $N=1000$, square box with $L_x=L_y=57.74$, alignment strength $\kap=40$, alignment range $\sig=1.25$, and particle density $\bar\rho= N/(L_x L_y)=0.3$ associated with a packing fraction $\phi \simeq 0.24$.

We model excluded volume via a short-ranged repulsive Weeks-Chandler-Andersen (WCA) potential
\begin{align}
U_{\text{WCA}}(r) =
    \begin{cases}
        4\epsilon \left[\left(\frac{a}{r}\right)^{12}-\left(\frac{a}{r}\right)^6\right]+\epsilon ~, & r \leq 2^{1/6}\\
        0 ~, & r > 2^{1/6}
    \end{cases} 
\label{su_eq:wca}
\end{align}
with $\epsilon=100$. The force acting on the $k$th particle is given by $\vec F_k=-\nabla_k U_\text{WCA}$.

To facilitate the study of traveling bands for different parameter values, we promote their formation by selecting appropriate initial configurations. Note that in case of a square simulation box ($L_x=L_y$), the system is naturally symmetric in both spatial directions, generally allowing bands to travel in $\pm x$ as well as $\pm y$-direction (see Supplemental Video 5). To bias the band along the positive $x$-axis, we set half of the initial orientations $\vhi_k = -2\pi x_k/L_x$ with $x$-coordinate of the $k$th particle position $x_k\in[-L_x/2, L_x/2]$ uniformly distributed, while the other half orientations are chosen randomly. This, assuming $\om_0>0$, reliably produces traveling bands propagating in positive $x$-direction. 

The system is translationally invariant along $y$-direction and we bin particles along $x$. Since bands are propagating, we shift the coordinate system, i.e., change into a comoving reference frame such that the bin with vanishing mean orientation along $y$, $\langle e_y \rangle_i =0$ (or closest to), is centered around the origin. Here $\langle \cdot \rangle_i$ denotes the average over particles in bin $i$. This represents the center of the band (cf. Figs.~1(b) and 2(d) of the main text).

\section{Hydrodynamic equations}

\subsection{Derivation}

In this section, we show how to derive the effective hydrodynamic equations of the main text from the microscopic model described by the Langevin equations for particles $k\in \lbrace 1,N \rbrace$
\begin{equation}
  \dot\x_k = v_0\vec e_k - \mu_0\vec F_k; \qquad \dot\vhi_k = \om_0 + \frac{\kap}{\pi\sig^2}\sum_{l\in n_k}\sin(\vhi_l-\vhi_k) + \sqrt{2/\tau}\eta_k,
  \label{sup_eq:motion}
\end{equation}
with Gaussian white noise $\eta_k$ in the orientation and alignment interaction strength given by $\kap$. The number of neighboring particles within interaction range $\sig$ is denoted by $n_k$. As a first step, we borrow a technique from the mean-field solution of the Kuramoto model~\cite{kuramoto75,acebron05} and eliminate the summation in the alignment interaction by defining an effective local angle $\theta_k$ through the relation
\begin{equation}
  \sum_{l\in n_k} e^{\im\vhi_l} = |n_k|r_k e^{\im\theta_k},
  \label{eq:trick}
\end{equation}  
with local phase coherence strength $0 \leq r_k \leq 1$. Applying Euler's formula and further replacing the (average) number of neighbors $|n_k|\approx \rho(\x_k)\pi\sig^2$ through the local density $\rho(\x_k)$ in close proximity to the $k$th particle, we get
\begin{equation}
  \dot\vhi_k = \om_0 + \kap\rho(\x_k)r_k\sin(\theta_k-\vhi_k) + \sqrt{2/\tau}\eta_k.
\end{equation}  
The corresponding Smoluchowski equation for the joint distribution $\psi(\x,\vhi;t)$ to find a single tagged particle at position $\vec r$ with orientation $\vec e$ reads
\begin{equation}
  \partial_t\psi = -\nabla\cdot(v_0\vec e\psi) - \pd{}{\vhi}[\om_0\psi+\kap\rho r\sin(\theta-\vhi)\psi] + \frac{1}{\tau}\pd{^2\psi}{\vhi^2},
  \label{sup_eq:smol_eq}
\end{equation}
with ``quenched'' fields $\theta(\vec r)$ and $r(\vec r)$. 

The effective hydrodynamic equations for the particle density and polarization 
\begin{equation}
  \rho(\x,t) = \IInt{\vhi}{0}{2\pi}\psi(\x,\vhi;t); \qquad \vec p(\x,t) = \IInt{\vhi}{0}{2\pi}\vec e\psi(\x,\vhi;t),
\end{equation}
take the form 
\begin{gather}
  \partial_t\rho=-\nabla\cdot(v_0\vec p); \label{sup_eq:evolution_den} \\
  \partial_t\vec p = -\nabla\cdot\left(\frac{1}{2} v_0\rho\id+v_0 \vec Q\right) + \om_0\vec R\cdot\vec p + \kap\rho r \left(\frac{1}{2}\rho\id - \vec Q\right)\cdot\vec n - \frac{1}{\tau}\vec p.
  \label{sup_eq:evolution_pol_interm}
\end{gather}
Here $\vec Q(\x,t)=\Int{\vhi}(\vec e\vec e-\id/2)\psi$ is the nematic tensor, the matrix $\vec R$ describes a counter-clockwise rotation by $\pi/2$, and we define the vector $\vec n \equiv (\cos\theta,\sin\theta)^\top$ to write $\sin(\theta-\vhi) = \vec n\cdot \vec R \cdot \vec e$.

\begin{figure}
  \centering
  \includegraphics{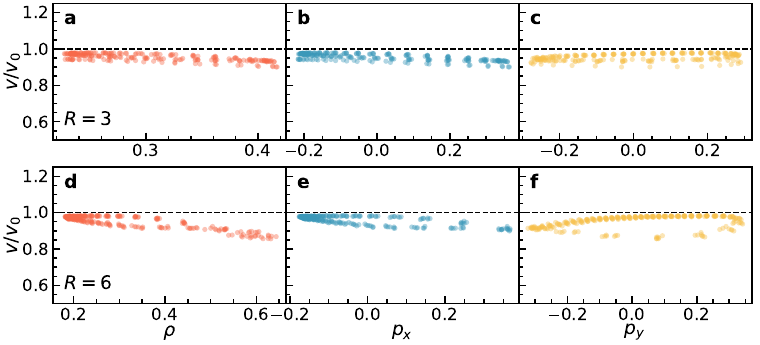}
  \caption{\textbf{Effective speed is approximately constant.} (a-c)~Ratio of effective speed $v$ and the bare propulsion speed $v_0$ as a function of (a)~the local density $\rho$, (b)~the $x$-component of the polarization density $p_x$, and (c)~the $y$-component of the polarization density $p_y$ for orbital radius $R=3$. (d-f)~Equivalent plots for intermediate radius $R=6$. Dashed black lines are a guide to the eye, highlighting the fact that $v\approx v_0$. Note that the $y$-axis starts at a value of $v/v_0=0.5$.}
  \label{fig:speeds}
\end{figure}

To proceed, we relate the fields $\theta$ and $r$ to the polarization. We multiply Eq.~\eqref{eq:trick} by the joint probability of orientations fixing particle $k$ and integrate out the orientations,
\begin{equation}
  \sum_{l\in n_k} \IInt{\vhi}{0}{2\pi}e^{\im \vhi} \psi(\x_l,\vhi) \approx |n_k|[p_x(\x_k)+\im p_y(\x_k)] = |n_k|r(\x_k)e^{\im\theta(\x_k)}\rho(\x_k).
\end{equation}
For the second step, we assume that interactions are short-ranged, $\psi(\x_l,\vhi)\approx\psi(\x_k,\vhi)$. As closure, we thus approximate $\vec p\approx r\vec n\rho$ and assume $\vec Q \approx 0$, i.e., the considered timescale is sufficiently long such that the nematic order has (almost) fully decayed. Plugging this into Eq.~\eqref{sup_eq:evolution_pol_interm}, we obtain the non-linear evolution equation for the polarization
\begin{equation}
  \partial_t\vec p = -\frac{v_0}{2} \nabla \rho + \om_0\vec R\cdot\vec p + \left(\frac{\kap}{2}\rho - \frac{1}{\tau}\right)\vec p.
  \label{sup_eq:evolution_pol}
\end{equation}
In principle, the excluded-volume interactions could renormalize the linear speed. However, the ratio of effective particle speeds $v$ measured in simulations (the averaged projection of particle displacements onto particle orientations per unit of time) to the bare speed $v_0$, plotted in Fig.~\ref{fig:speeds}, shows that assuming constant particle speeds---independent of local density and polarization---is a well-justified approximation.

Motivated by the behavior displayed in simulations, we seek an evolution equation that allows for (non-zero) periodic solutions---especially if alignment interactions overshadow rotational diffusion ($\chi = \kap\bar \rho/2 - \tau^{-1}> 0$)---which is not the case for Eq.~\eqref{sup_eq:evolution_pol}. To see this, consider the homogeneous case $\rho = \bar \rho$. It is straightforward to find solutions that, depending on the sign of $\chi$, either decay ($\chi < 0$) or grow ($\chi > 0$) exponentially. Therefore, we include a correction term at the lowest possible order of $\vec p$,
\begin{equation}
  \partial_t\vec p = -\frac{v_0}{2} \nabla \rho + \om_0\vec R\cdot\vec p + \left(\frac{\kap}{2}\rho - \frac{1}{\tau} - \al |\vec p|^2\right)\vec p,
  \label{sup_eq:evolution_pol_alpha}
\end{equation}
with a phenomenological parameter $\al>0$ that is treated as a fit parameter. We posit that this term arises through interactions (and the closure relation for the nematic tensor) and prevents uncapped exponential growth of the amplitude in the case of $\chi>0$ as considered in this work.

\subsection{Comment on alternative hydrodynamic equations}

In a recent publication, Sesé-Sansa and coworkers also derive effective hydrodynamic equations for a system of self-propelled chiral particles that experience generic torques~\cite{sese-sansa22} related to the microscopic model that we study here. In Ref.~\cite{sese-sansa22}, a different closure scheme is employed to the particle torques---reminiscent of the force closure introduced by Bialke \emph{et al.}~\cite{bialke13}---leading to an effective density-dependent angular speed $\om(\rho)$. To arrive at this result, all two-body terms describing particle interactions are grouped into a single scalar coefficient, interpreted as a rotational friction coefficient. However, to eventually arrive at hydrodynamic equations one has to assume a (or close to) homogeneous state, for which this scalar coefficient is simply a constant. In contrast, our derivation does not call for such a torque closure since the appropriate treatment of alignment interactions follows naturally from the analogies drawn from the well-studied Kuramoto model. Thus, we are able to derive hydrodynamic equations that hold for inhomogeneous density profiles, which is essential for our investigation of traveling bands.

\subsection{Relation to Vicsek flocks}

These traveling bands are fundamentally different from polarized flocks of non-chiral particles, which also displays density ``bands'' (typically referred to as flocks) that travel at constant speed. It is now understood that, in the Vicsek model~\cite{solon15a}, these bands correspond to the coexistence of active gas and polar liquid. In contrast, the polarization following from Eq.~\eqref{sup_eq:evolution_pol_alpha} is not constant within the band, but changes along the direction of propagation. Additionally, the propulsion mechanism is more complex than in polar bands where, up to small perturbations, the band speed is equal to the particle propulsion speed $v_0$.

It is instructive to see how the evolution equation of Ref.~\citenum{solon15a}
\begin{equation}
  \partial_t \vec p + {\color{blue}\zeta (\vec p \cdot \nabla) \vec p} = -\frac{v_0}{2} \nabla \rho + \left(\frac{\kap}{2}\rho - \frac{1}{\tau} - \al |\vec p|^2\right)\vec p + {\color{blue}D_0 \nabla^2 \vec p}.
  \label{sup_eq:vicsek-like}
\end{equation}
differs from our result (beyond the trivial $\om_0=0$). First, we need to introduce an effective translational diffusion term $D_0\nabla^2\vec p$ with diffusion coefficient $D_0$. Second, and more importantly, we need to include the non-linear self-advection term $\zeta (\vec p \cdot \nabla) \vec p$ with another coefficient $\zeta$ stemming from the closure at nematic order (see, e.g.~\cite{farrell12,jayaram20}). Following the analysis of Ref.~\citenum{solon15a}, one finds inhomogeneous polar excitations by mapping Eq.~\eqref{sup_eq:vicsek-like} onto an effective Newtonian system. Exploration of phase space reveals three permitted solutions:~(i) periodic orbits corresponding to periodic waves in density and polarization, (ii) homoclinic cycles representing isolated solitonic bands, and (iii) heteroclinic cycles indicating phase separation into polar-liquid droplet and isotropic gas. Lastly, allowing $\alpha = \alpha(\rho)$ to explicitly depend on density, one recovers the correct polarization scaling, i.e., its saturation as function of density, as exhibited by a polar ordered phase of the Vicsek model.

\section{Linear stability of the synchronized state}

The synchronized state has a spatially uniform polarization
\begin{equation}
  \vec p^{(\text{s})}(t) = \sqrt{\chi/\al}\left(
    \begin{array}{c}
      \cos\om_0t \\ \sin\om_0t      
    \end{array}\right)
\end{equation}
together with a uniform density $\bar\rho$. We now consider small perturbations to Eqs.~\eqref{sup_eq:evolution_den} and \eqref{sup_eq:evolution_pol_alpha} through $\rho=\bar\rho+\delta\rho$ and $\vec p=\vec p^\text{(\text{s})}+\delta\vec p$, leading to the linearized equations
\begin{gather}
  \partial_t\delta\rho = -v_0\nabla\cdot\delta\vec p \\
  \partial_t\delta\vec p = -\frac{v_0}{2}\nabla\delta\rho + \om_0\vec R\cdot\delta\vec p + \left(\frac{\kap}{2}\delta\rho - 2\al\vec p^{(\text s)}\cdot\delta\vec p\right)\vec p^{(\text s)}.
  \label{sup_eq:pol_linearized}
\end{gather}
The last term of the second equation describes the relaxation (or growth) of the perturbation $\delta\vec p$ along the direction of $\vec p^{(\text{s})}$. We make the ansatz $\delta\vec p(\x,t)=q_\parallel(\x,t)\vec p^{(\text{s})}(t)$ leading to
\begin{gather}
  \partial_t\delta\rho = -v_0\vec p^{(\text{s})}\cdot\nabla q_\parallel; \\
  \frac{\chi}{\al}\partial_t q_\parallel = -\frac{v_0}{2}\vec p^{(\text{s})}\cdot\nabla\delta\rho + \left(\frac{\kap}{2}\delta\rho - 2\chi q_\parallel\right)\frac{\chi}{\al}
\end{gather}
after projecting Eq.~\eqref{sup_eq:pol_linearized} onto $\vec p^{(\text{s})}$. The evolution of $q_\parallel$ is determined by the competition between alignment $\kap$ and relaxation $\chi$ (which itself depends on $\kap$).

\begin{figure}[t]
  \centering
  \includegraphics{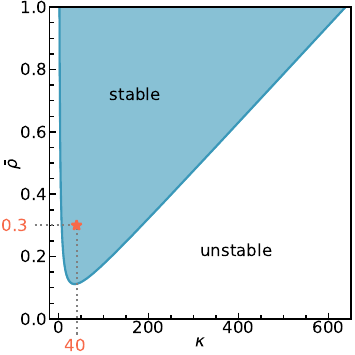}
  \caption{\textbf{Linear stability.} The synchornized state is linearly stable for values of global density $\bar\rho$ and alignment strength $\kap$ that fall in the shaded region. The orange star ($\bar\rho=0.3$ and $\kap=40$) corresponds to the parameter values we study in simulations. We set the fit parameter $\al=160$ in accordance with the fitted current data in Fig.~(3) of the main text.}
  \label{fig:sync}
\end{figure}

We now go into a rotating frame through $x=\xi\cos\om_0t$ and $y=\xi\sin\om_0t$, so that $\partial_\xi (\cdot)=\cos\om_0t\partial_x (\cdot)+\sin\om_0t\partial_y(\cdot)$ and thus
\begin{gather}
  \partial_t\delta\rho = -v_0\sqrt{\chi/\al}\partial_\xi q_\parallel; \\
  \frac{\chi}{\al}\partial_t q_\parallel = -\frac{v_0}{2}\sqrt{\chi/\al}\partial_\xi\delta\rho + \left(\frac{\kap}{2}\delta\rho - 2\chi q_\parallel\right)\frac{\chi}{\al}.
\end{gather}
We insert the ansatz $\{\delta\rho,q_\parallel\}\to\{\delta\rho(t),q_\parallel(t)\}e^{\im k\xi}$ with time-independent $\xi$, which leads to the coupled evolution equations
\begin{equation}
  \partial_t\left(\begin{array}{c}
    \delta\rho \\ q_\parallel
  \end{array}\right) =
  \left(\begin{array}{cc}
    0 & -\im v_0\sqrt{\chi/\al}k \\
    -\im\tfrac{1}{2}v_0\sqrt{\al/\chi}k + \tfrac{\kap}{2} & -2\chi
  \end{array}\right)
  \left(\begin{array}{c}
    \delta\rho \\ q_\parallel
  \end{array}\right)
\end{equation}
for the amplitudes. The eigenvalues of the matrix, which correspond to the growth rates, read
\begin{equation}
  \sig_\pm(k) = -\chi \pm \sqrt{\chi^2 - \im\frac{1}{2}v_0\kap\sqrt{\chi/\al}k - \frac{1}{2}v_0^2k^2}.
\end{equation}
The non-zero imaginary part of $\sig_\pm$ implies oscillations. The real part of $\sig_-$ is always negative while an expansion for $\sig_+$ leads to
\begin{equation}
  \sig_+(k) \approx -\chi + \chi\left[ 1 - \im\frac{1}{4\chi^2}v_0\kap\sqrt{\chi/\al}k + \frac{1}{32}\left(\frac{\kap^2}{\al\chi^3}-\frac{8}{\chi^2}\right)v_0^2k^2 + \cdots \right].
\end{equation}
Hence, $\sig_+(k\to0)=0$ as required for dynamics that conserve density. The system becomes unstable at $\kap_\text{c}^2=\al(4\kap_\text{c}\bar\rho-8\tau^{-1})$ with critical alignment parameter
\begin{equation*}
  \kap_\text{c}^\pm = 2\al\bar\rho \pm 2\sqrt{\al^2\bar\rho^2-2\al\tau^{-1}},
\end{equation*}
cf. Fig.~\ref{fig:sync}. The synchronized phase thus becomes linearly unstable both for small and large $\kap$. For small $\kap$, rotational diffusion dominates, whereas for large $\kap$, local alignment suppresses global synchronization.

\section{Perturbative solution}

In the comoving frame of reference ($x \to x-ct$), the polarization follows the evolution equation
\begin{equation}
  -c\partial_x \vec p = -\frac{v_0}{2} \partial_x \rho \hat{\vec e}_x + \om_0 \vec R \cdot \vec p + \left( \frac{\kap}{2}\rho- \frac{1}{\tau} -\al |\vec p|^2 \right)\vec p.
  \label{sup_eq:pol_comoving}
\end{equation}
Here we have replaced $\partial_t \to -c\partial_x$ in Eq.~\eqref{sup_eq:evolution_pol_alpha} and neglected all $y$-dependencies due to translational invariance. Assuming $\eps = v_0/c \ll 1$, which is equivalent to small $qR\ll 1$ ($q$ will become the wave vector magnitude of the periodic solution), we expand $\vec p=\vec p^{(0)}+\eps\vec p^{(1)}+\cdots$ and perturbatively solve Eq.~\eqref{sup_eq:pol_comoving} up to linear order in $\eps$. With the lowest order solution given by [Eq.~(6) in the main text]
\begin{equation}
  \vec p^{(0)}(x;q) = \sqrt{\chi_0/\al}\left(\begin{array}{c}
    \cos qx \\ -\sin qx
  \end{array}\right),
  \label{sup_eq:p0:sol}
\end{equation}
where $\chi_0 = \kap\rho_0/2 -\tau^{-1}$ (without setting $\rho_0=\bar \rho$ yet), we now provide more details on the solution to linear order in $\eps$.

We have to solve 
\begin{equation}
  -c \partial_x \vec p^{(1)} = \om_0 \vec R \cdot \vec p^{(1)} + \frac{\kap}{2} p_x^{(0)} \vec p^{(0)} - 2\al \left(\vec p^{(0)} \cdot \vec p^{(1)} \right) \vec p^{(0)}.
\end{equation}
Notably, the first term on the right-hand side of Eq.~\eqref{sup_eq:pol_comoving} is $O(\eps^2)$. Plugging in the lowest order solution, Eq.~\eqref{sup_eq:p0:sol}, and employing the periodic ansatz $\vec p^{(1)}(x;q) = a(x) (\cos qx,-\sin qx)^\top$, the oscillation amplitude is determined by the ordinary differential equation
\begin{equation}
a'(x) = -\frac{\kap \chi_0}{2c\al} \cos(qx) + \frac{2\chi_0}{c} a(x)
\end{equation}  
with solution 
\begin{equation}
  a(x) = \frac{\kap \chi_0}{\al(4\chi_0^2+ \om_0^2)} \left( \chi_0 \cos(q x) - \frac{\om_0}{2} \sin(q x)  \right)
\end{equation}  
after enforcing periodic boundary conditions and $q=q_n=2\pi n/L_x$. This gives Eq.~(9) from the main text.

\begin{figure}[t]
  \centering
  \includegraphics{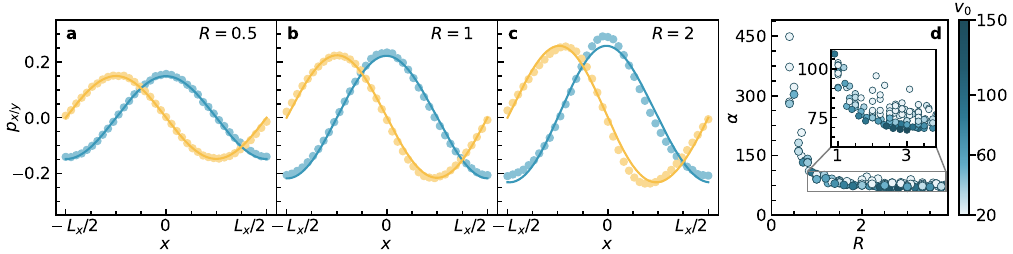}
  \caption{\textbf{Fitted solution and fit parameter.} (a-c)~Simulated data (dots), fitted by the perturbative solution of the hydrodynamic equations to first order in $\eps$ for $R\in\lbrace 0.5, 1, 2\rbrace$ (solid lines). Yellow displays the $y$-component of the polarization and blue the $x$-component. The box size is $L_x=57.74$. (d)~Fitted non-linear coefficient $\al$ for moderately small $R \in [0,4)$. Face colors represent the self-propulsion speeds $v_0$.}
  \label{fig:fits}
\end{figure}

Fitting the solution to the simulated data, with single fit parameter $\al$, we can see excellent agreement for reasonably small $R\leq 2$, as displayed in Fig.~\ref{fig:fits}(a-c). The corresponding $\al$-values are shown in Fig.~\ref{fig:fits}(d), where face colors represent the self-propulsion speed $v_0$. The small amplitude of the simulated polarization for $R\lesssim 1$ results in large $\al$. Upon increasing R, the value of $\al$ decreases and eventually saturates to a constant $\al \approx 65$. Notably, close inspection of the marker colors in the inset of Fig.~\ref{fig:fits}(d) hints towards a dependence $\alpha(v_0)$, where large $v_0$ tend to result in lower values for $\alpha$ and vice versa.

\subsection{Density conservation}

Due to translational invariance along the $y$-axis, we saw that the density profile in the comoving frame is given by $\rho(x) = \rho_0 + \eps p_x(x)$, and thus, apart from a constant off-set, determined by the $x$-component of the polarization. To compute the constant contribution $\rho_0$, we enforce the condition
\begin{equation}
  \IInt{x}{0}{L_x} \rho(x) = L_x \bar \rho,
\end{equation}
to guarantee mass conservation. At lowest order we find that $\rho_0 = \bar \rho$, due to the symmetry of $p_x^{(0)}$. This \emph{a priori} no longer holds at first order, since integration results in $\rho_0 = \bar \rho - \eps^2 \kap\chi^2/[2\al(4\chi^2 + \om_0^2)]$. However, the correction is $O(\eps^2)$ and hence negligible for the considered small-$R$ regime. Consequently, setting $\rho_0 = \bar \rho$ and accordingly $\chi_0=\chi$ is well justified whilst discussing the perturbative solution and its implications.

\section{Mass transport}

\begin{figure}[b!]
  \centering
  \includegraphics{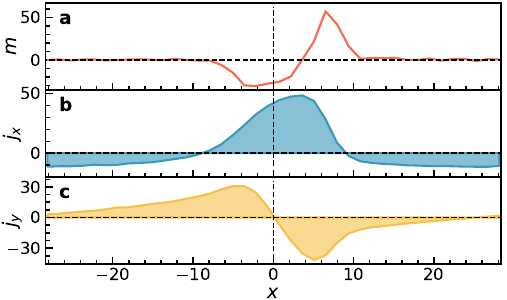}
  \caption{\textbf{Torque and current profile.} (a)~Average torque $m(x)$ experienced by particles in a bin calculated according to Eq.~\eqref{sup_eq:torque}. (b)~$x$- and (c)~$y$-component of the currents per bin $j_{x/y}(x)$. The displayed data corresponds to a simulation with $R=10$. Dashed black lines indicate the coordinate system of the comoving frame.}
  \label{fig:cur}
\end{figure}

The onset of a dense traveling band introduces an effective alignment field around the band center ($x=0$). We calculate binned torques 
\begin{equation}
  m(x_i) = \left\langle \frac{\kap}{\pi\sig^2}\sum_{l\in n_k}\sin(\vhi_l-\vhi_k)\right\rangle_{i}
   \quad \text{such that} \quad \sum_{i=1}^{N_\mathrm{b}} m(x_i)=0,
    \label{sup_eq:torque}
\end{equation}
where $\langle \cdot \rangle_i$ denotes the average over all particles in bin $i \in \lbrace 1,\ldots,N_\mathrm{b} \rbrace$ centered around position $x_i$. Figure~\ref{fig:cur}(a) shows that the experienced torques undergo a sign change while traversing the band and vanishes outside the band. Intrinsic rotations of particles assimilated at the band front are amplified, promoting alignment along $x$ as dictated by the dense band, whereas rotations of departing particles are suppressed. Consequently, the overall time particles spend pointing along $x$ is increased and individual particles do not return to their initial positions after one period. As discussed in the main text, this underlies the emergence of a net particle current along the band propagation direction, which is substantiated by looking at the profile of the $x$-component of time-averaged currents per bin $j_x(x)$, displayed in Fig.~\ref{fig:cur}(b). Clearly, particles (especially those within the dense region) show a preference to move along positive $x$-direction. Furthermore, since the accelerated rotation at the band front reduces the duration of particles facing downwards, one additionally observes mass transport in positive $y$-direction. This, although less obvious, can be deduced from Fig.~\ref{fig:cur}(c), where the area enclosed by $j_y(x)$ and the $x$-axis is positive.

In the laboratory frame, currents through perpendicular cross sections---averaged over one period $\tau_1=L_x/c$---read
\begin{equation}
  J_x = \frac{1}{\tau_1}\IInt{t}{0}{\tau_1}\IInt{y}{0}{L_y} v_0p_x(x,t); \quad 
  J_y = \frac{1}{\tau_1}\IInt{t}{0}{\tau_1}\IInt{x}{0}{L_x} v_0p_y(x,t).
\end{equation}
Going to the comoving frame, i.e., $x \to x-ct$, and subsequently replacing $c \IInt{t}{0}{\tau_1} \to \IInt{x}{0}{L_x}$, we obtain
\begin{equation}
  J_x = v_0\frac{L_y}{L_x}\IInt{x}{0}{L_x} p_x(x); \qquad
  J_y = v_0\IInt{x}{0}{L_x} p_y(x).
\end{equation}
Hence, both (integrated) currents are independent of the position of the cross section, as one would assume for a periodic system.
The calculated currents are reported in Eq.~(11) of the main text and show a quadratic dependence on self-propulsion speed $v_0$. Recalling the above explanation, the dependence should come as no surprise, since effects of the alignment imbalance are increasingly amplified (suppressed) the faster (slower) particles move.

\begin{figure}[t]
  \centering
  \includegraphics{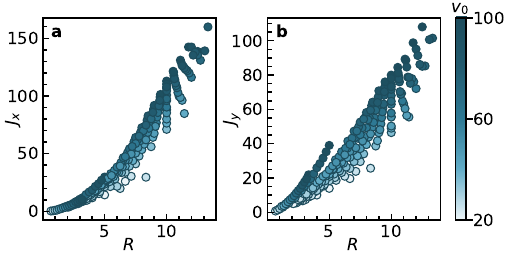}
  \caption{\textbf{Unscaled particle currents.} (a) $J_x$ and (b) $J_y$ from simulations, as a function of orbital radius $R$. Colors denote self-propulsion speeds $v_0 \in [20,150]$. Both current components show a clear $v_0$ dependence, where higher (lower) speeds amplify (reduce) the amplitude of mass transport.}
  \label{fig:cur_unscaled}
\end{figure}

Lastly, in Fig.~\ref{fig:cur_unscaled}, we plot the currents $J_x=\Delta x/(L_x\tau)$ and $J_y=\Delta y/(L_y\tau)$ measured in simulations where $(\Delta x,\Delta y)^\top$ is the average displacement of particles in time $\tau$. Inspection of marker colors evidently displays the dependence on the self-propulsion speed $v_0$, where, as argued, large (small) speeds increase (reduce) the current magnitude. 

\section{Breaking of band} 
\begin{figure}[h!]
  \centering
  \includegraphics{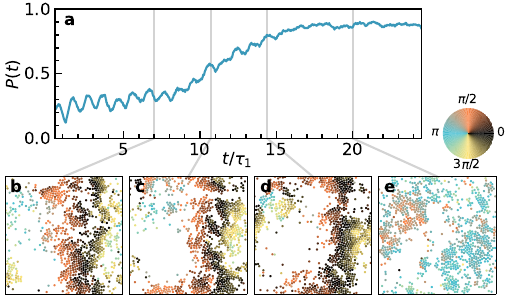}
  \caption{\textbf{Gradual breakdown of traveling bands.} (a)~Order parameter $P(t)$ for $R=5$ showing a gradual increase over $10$ to $15$ rotation periods $\tau_1$. (b-e)~Snapshots of the system showing the band's transition to the synchronized state.}
  \label{fig:break}
\end{figure}

To characterize the traveling band phase in simulations, we turn towards the order parameter 
\begin{align}
  P(t) \equiv \frac{1}{N} \left| \sum_{k=1}^N \vec e_k(t) \right|, 
\end{align}
describing the average global orientation. In case of traveling bands, this parameter should be close to zero, since no orientation is globally preferred. In the synchronized state, however, particles point along the same direction and $P\to 1$. We consider the system to be a traveling band if $P<0.5$ for a time span of $\Delta t=5$.

Although bands are generally stable in the small-$R$ regime, occasionally rare fluctuations can destabilize traveling bands and initiate a transition towards the synchronized cluster phase. However, unlike the rapid breakdown displayed for large $R$ [cf.~Fig.~4 of the main text], wherein the synchronized phase is reached within roughly $2\tau_1$, for small $R$, the system dwells in transient states for an extended period of time ($10\tau_1$ to $15\tau_1$) before eventually reaching the stable synchronized phase (see Fig.~\ref{fig:break}).

\end{document}